\documentclass[journal]{IEEEtran}
\usepackage[noadjust]{cite}

\usepackage{caption}
\usepackage{subcaption}
\usepackage[export]{adjustbox}
\usepackage{graphicx}
\graphicspath{{figures/}}

\usepackage{siunitx}
\usepackage{etoolbox}

\usepackage{float}

\ifCLASSINFOpdf

\else

\fi

\usepackage{amsmath,amsfonts,amssymb}
\usepackage[table,xcdraw]{xcolor}
\usepackage{multirow}
\begin{document}

\title{Predictive Display with Perspective Projection of Surroundings in Vehicle Teleoperation to Account Time-delays}


\author{Jai~Prakash,~Michele~Vignati,~Daniele~Vignarca,~Edoardo~Sabbioni,~and~Federico~Cheli

\thanks{The authors belong to the Department of Mechanical Engineering, Politecnico Di Milano, Italy (e-mail:
jai.prakash@polimi.it; michele.vignati@polimi.it; daniele.vignarca@polimi.it; edoardo.sabbioni@polimi.it; federico.cheli@polimi.it)
}%

\thanks{This work was financed by project FESR 2014-2020 ID 242092 - TEINVEIN CUP E96D17000110009}
}


\maketitle


\begin{abstract}

Teleoperation provides human operator sophisticated perceptual and cognitive skills into an over-the-network control loop. It gives hope of addressing some challenges related to vehicular autonomy which is based on artificial intelligence by providing a backup plan. Variable network time-delays in data-transmission is the major problem in teleoperating a vehicle. On 4G network, variability of these delays is high (70-150 ms ping). Due to this, both video streaming and driving commands encounter variable time-delay. This paper presents an approach of providing the human-operator a forecasted video stream which replicates future perspective of vehicle’s field of view accounting the delay present in the network. Regarding the image transformation, perspective projection technique is combined with correction given by smith predictor in the control loop. This image transformation accounts current time-delay and tries to address both issues, time-delays as well as its variability. For experiment sake, only frontward field of view is forecasted. Performance is evaluated by performing online vehicle teleoperation on street edge-case manoeuvres and later comparing the path deviation with and without perspective projection.

\end{abstract}

\begin{figure}[h]
    \centering
    \includegraphics[width=0.4\textwidth]{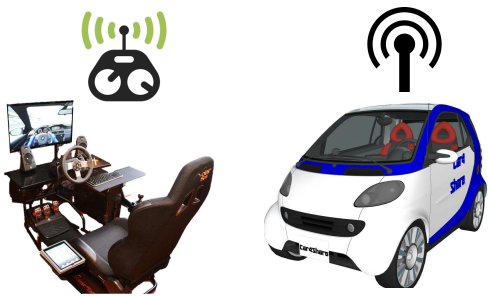}
    \caption{Vehicle tele-operation}
    \label{fig:x vehicleTeleoperation}
\end{figure}

%
\section{Introduction}
%
%
%
%
\IEEEPARstart{T}{eleoperation} indicates operation of a system from a distance. It means that there is no direct interference between human operator and teleoperated object. In vehicle teleoperation, teleoperated object is the vehicle (figure \ref{fig:x vehicleTeleoperation}). The human-operator controls the vehicle from the control station while the car is on the road. Control station 
is a fixed control-center that provides facilities for human interaction. It consists of display screens, speakers, and steering wheel/pedals control. \textcolor{black}{Human-operator steers the steering joystick \cite{Graf, Gnatzig}, then the commands reach the vehicle through a data communication protocol}. Later, the commands get actuated in the vehicle through the actuators installed inside the vehicle.

The communication protocol can either be wired or wireless. To avail maximum benefits of teleoperation, this work aims wireless communication based on 4G LTE wireless broadband communication which is widely available across the world. This makes it a compelling choice to be used as data communication medium. With 4G, a vehicle can be operated miles away from its real location. 
\textcolor{black}{Its potential applications could be remote last mile delivery of rental/shared vehicles, avoiding driver presence in dangerous areas, human assistance in case of fallback of autonomous vehicles, and valet parking etc.} Because of many reasons, the driving experience of the human operator would not be exactly same as compare to driving the vehicle from inside of it. First, human-operator is not able to feel vehicle acceleration while sitting on a seat inside the control station. Second, although display screens try to emulate windshield view, display screens are usually smaller than windshield. Moreover, in cases where camera is mounted outside the vehicle on its roof for better sensing, human operator observes different perspective as compare to perspective from inside of the vehicle. Altogether, visual input to human operator is different than visual input to a real vehicle driver. Third, is the absence of real haptic feedback. Haptic feedback are steering feedback force, pedal press force. Fourth, and most critical aspect in teleoperation, is time-delay or latency. Time-delay is defined as the time passing between the user’s input and its displayed response \cite{Watson1998}. For human in-the-loop control systems, time-delay has been considered to affect performance factors. The accuracy of control actions deteriorates because of operator’s inability to visualize or predict the outcome of his control actions. Humans can adapt to time-delays in control systems \cite{Cunningham2001}, however, this adaptation depends on human operator ability to predict the outcome of his control actions \cite{Sheridan1993}, and the extent of this adaptation is dependent upon the characteristics of the time-delay (e.g., magnitude and variability)\textcolor{black}{ \cite{Davis2010}}.


Studies on human performance in virtual environment show that people are generally able to detect latency as low as 10–20 ms \cite{Ellis2004}. In a simulated driving task, the driver’s vehicle control is found to be significantly degraded with a latency of even 170 ms \cite{Frank1988}.

\begin{figure}[h]
    \centering
    \includegraphics[width=0.4\textwidth]{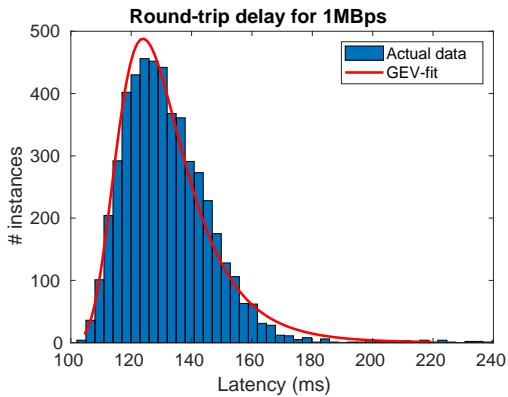}
    \caption{Round trip time-delay data}
    \label{fig:x roundTripLatency}
\end{figure}

Time-delay components in teleoperation are camera frame capture delay (exposure delay), data communication delay, image-processing delay, human-operator response time, and vehicle actuation delay. Camera's capture delay with a general purpose USB3.0 camera for $672$x$376$ image frame is $\sim$70ms. It can be considered constant for constant illuminance of ambient light. \textcolor{black}{Due to independent treatment of pixels (section II-B-2),} image-processing delay is constant for constant image resolution. Human-operator response time is present in both teleoperation and real vehicle drive. Data communication delay is the bottleneck in time-delay cycle. 
During vehicle teleoperation, vehicle streams images to the control-station in the form of jpeg images and the control-station transmits driving commands to the vehicle as better described in the following. Bandwidth consumption during this data communication is around 1 MBps.
Figure \ref{fig:x roundTripLatency} shows the round-trip delay observed while performing vehicle teleoperation.
Here the data corresponds to 1000 image frames and corresponding 1000 driving commands. 
This test is performed in typical urban environment with the vehicle is connected to 4G mobile communication and control-station is connected to internet using wired LAN. Lastly, dynamics of vehicle actuation in non-aggressive driving can be considered fast and delay associated to it can be considered constant ($\sim20$ms).

Several mechanisms have been demonstrated to counter the effects of time-delay including mathematical predictors/filters, predictive displays, command displays, and observer based model. Mathematical filters/predictors are Kalman filter predictors and Smith predictors. Predictive and command displays try to counter the time-delay issue by providing immediate feedback to the operator through model representation. Command displays differ from predictive displays, as they require the remote system to be autonomous \cite{Lane2001}. Smith predictor approach \cite{Zheng2021416, Lee2017687, Kuzu201879, Kim20177221, Lee2018835, Batista201834} has the potential to mitigate the negative effect of the time delay. But it is highly dependent on the accuracy of the predictor model. T. Teng and P. Grant \cite{Teng}, uses online parameter estimation techniques for the predictor model, \textcolor{black}{to adapt changes in predictor model in real-time}.

R. Liu, D. Kwak \emph{et al}. \cite{Liu2017}, investigated two visual information display arrangement strategies: (i) presenting display frames to a remote driver as soon as the frames arrive, and (ii) smoothing the display by adding additional delay when necessary to the received frame to mitigate the delay variance. They demonstrated that in $2^{nd}$ arrangement, mitigates the negative effect of variable delay incurred by LTE networks. \textcolor{black}{In both strategies, operator is responsible to adapt to the delay and take maneuvering decisions by cognitively predicting vehicle position, which induces mental fatigue.}

An observer-based framework  \cite{Tandon2013, Xinyi2015, Zheng2019, Huba2017269}, in which model free approach is used to estimate the undelayed state of the teleoperator vehicle. Here, the closed-loop dynamics of the observer is based on a sliding surface. A blended prediction of the heading signal is considered by linearly combining the model-free prediction with a steering-model. Y. Zheng \emph{et al}. \cite{Zheng2020} found that the blended architecture offers improvement when compared to the purely model-free realization. \textcolor{black}{However, in human in the loop experiments, view at the predicted position is generated directly from the simulation environment and not by the image processing of the delayed image.}

\textcolor{black}{Using command displays approach}, Michael Fennel\cite{Fennel2021}, proposed an offline path follower where the operator “draws” a desired 2D path by walking in a large-scale haptic interface while a guiding force is exerted, which ensures that the generated path can later be accurately followed by a path tracking controller running offline on a remote robot. \textcolor{black}{However, in this strategy, operator is not actively controlling the task}.

Tito Chen \cite{tito2015}, proposed a safety concept for teleoperated vehicles, called free corridor, with which the path of an emergency braking of the vehicle is calculated continuously and visualized as augmented reality in the received images at the operator workstation. With this concept, it is the task of the human operator to continuously hold this path free while driving. In case of a communication loss, an emergency brake will be activated and the vehicle will brake along the before-visualized path. \textcolor{black}{However, in estimating the free corridor, the delay associated to the driving command from control-station to vehicle is not considered. Which may cause slight deviation in the predicted and actually realized vehicle trajectory.}

In a previous work \cite{Prakash2019}, the authors briefly introduce \emph{Perspective projection} technique. Perspective projection (PP) or perspective transformation is a linear projection where three dimensional objects are projected on an image plane \cite{hearn:1997}. There the authors present post-processing results i.e., data are collected with the vehicle and processed offline to validate the algorithm. The approach was found to be able to generate the new perspective of the forward displaced camera, taking input as the image captured at the previous position of the camera.

\textcolor{black}{In literature \cite{Zheng2021416, Lee2017687, Kuzu201879, Kim20177221, Lee2018835, Batista201834, Tandon2013, Xinyi2015, Zheng2019, Huba2017269, Zheng2020, tito2015}, position of the vehicle is estimated using predictive techniques and then either predicted position is overlayed on the delayed image using coloured markers
or field of view (FOV) at the predicted position is generated directly from the simulation environment by placing a virtual camera at predicted position (which is infeasible in reality). This paper utilizes PP to generate the FOV at the predicted position in reality
, which provides realistic perspective compare to using markers.} The undelayed states are predicted by the Smith-predictor in the control-station which employs a model of the vehicle. The predicted position corresponds to the position where the vehicle would receive driving commands back, corresponding to the input image it has sent before. \textcolor{black}{For initial performance check of PP in real-life vehicle teleoperation, only front camera is used in experiments.}


\textcolor{black}{
The core contributions of this paper are as follows:
\begin{itemize}
    \item Perspective Projection approach to augment the delayed image
    instead of using markers. It provides better realistic perception to the human operator. The technique is detailed in section II-B-2.
    \item To allow 4G transmission of big depth-map data, a logarithmic encoding is used to utilise JPEG compression, which is explained in section II-A-3.
\end{itemize}
}

\textcolor{black}{5G would decrease network latency, but latency would still be present in transmitting big data like images as compared to a simple ping. Apart of network latency, net latency consists of other factors such as camera exposure delay, data processing delay and actuator delay (see section II-B-1). Availability of 5G would further enchance performance of proposed approach by reducing network latency.
}

The paper is organized as follows. In section II, the modified teleoperation control loop employing the smith predictor is described. Then, the deployed architecture is presented with the explanation of its sub-elements. The sub-elements consists of the vehicle, sensing architecture, plant model for the smith predictor and PP. Section III presents the experiment procedure and shows the results of the experiment performing real-life vehicle teleoperation on street edge-cases. Section IV carries discussion on the respective results and discusses the benefits observed by employing PP. Section V concludes this work.
\section{Method}
\begin{figure}[ht]
\centering
\begin{subfigure}[b]{0.48\textwidth}
    \centering
    \includegraphics[width=\textwidth]{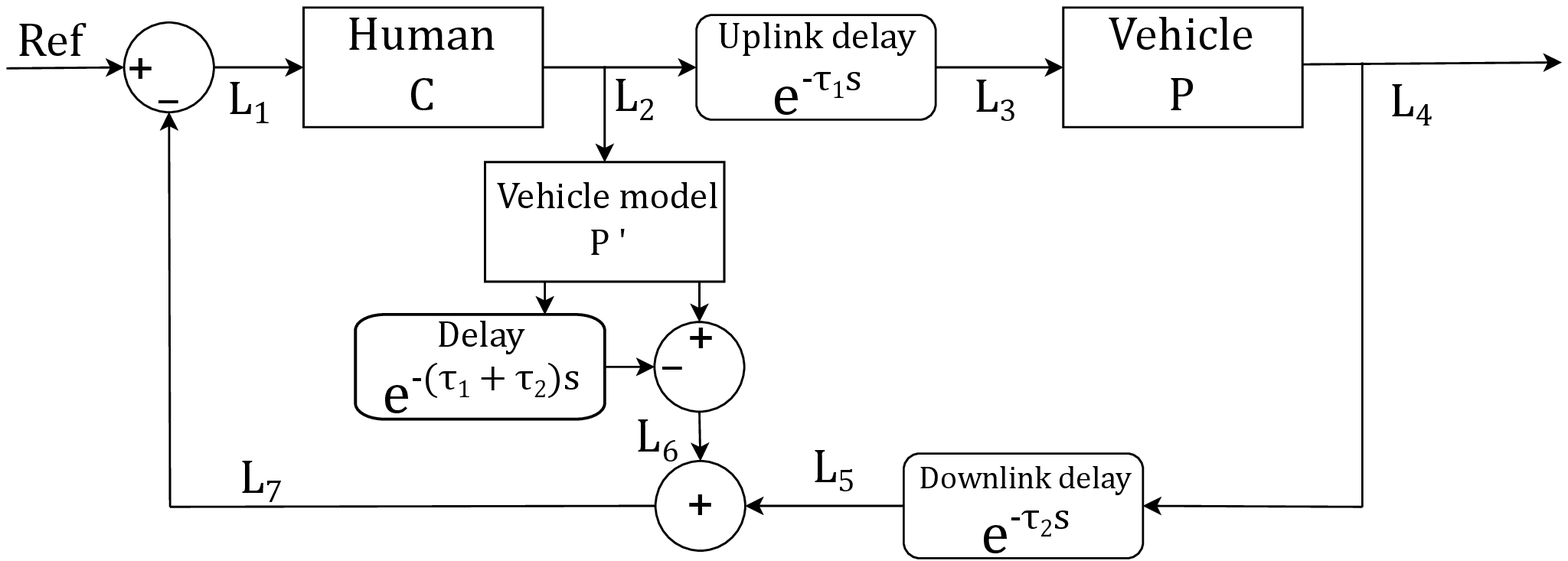}
    \caption{Smith predictor scheme}
    \label{fig:a smith}%
\end{subfigure}%
\vfill
\begin{subfigure}[b]{0.48\textwidth}
    \centering
    \includegraphics[width=\textwidth]{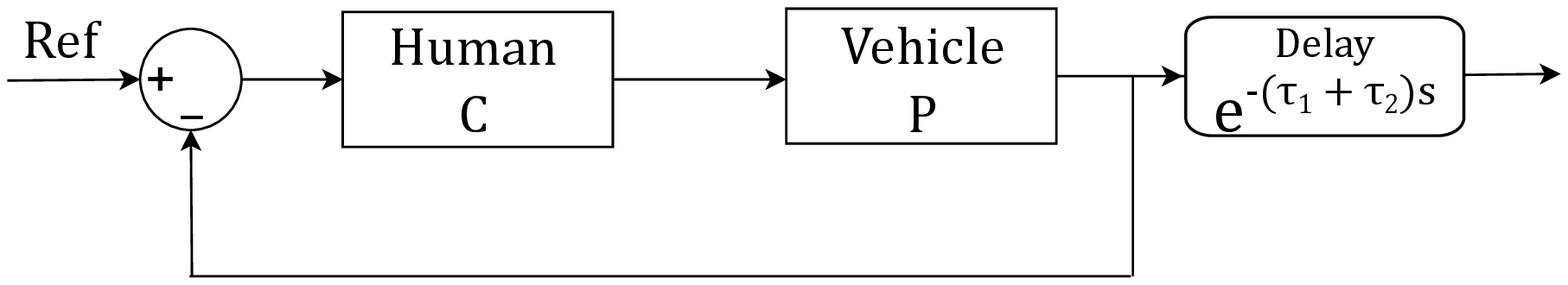}%
    \caption{Apparent closed loop system when P matches perfectly with P\rq  and delays are known}
    \label{fig:b smith}
\end{subfigure}
\caption{Smith predictor model - \textcolor{black}{a) Scheme, b) Apparent closed loop system}}
\label{smith predictor}
\end{figure}

The system model adopted for this vehicle teleoperation closely resembles to Smith Predictor model \cite{smith} as shown in figure \ref{smith predictor}. In Smith predictor setup, the control input \textcolor{black}{($L_2$)} is passed through a predictor model \textcolor{black}{$P'$} of the vehicle, which then passes through a transfer function \textcolor{black}{given by eq. \ref{eq1}}.
\begin{equation}
    TF = 1-e^{-\tau s}\label{eq1}
\end{equation} 
Where $\tau$ is the time delay between output of the controller and the respective feedback it receives from the plant in the control loop. Historically designed for classical control, the purpose of the Smith Predictor is to bypass the time delay in the observation, and transform the system into a pure forward delayed system \cite{Teng}. This is helpful for human-in-the-loop teleoperation, as it allows the human operator to not to wait for the feedback and provides the sense of controlling the vehicle in real-time.

In vehicle teleoperation, link $L_5$ in figure \ref{fig:a smith} corresponds to delayed vehicle states i.e., vehicle pose in the surrounding environment. In particular, link $L_5$ is the delayed image streaming transmitted by the vehicle and received by the control station. Link $L_6$ contains the supplement contribution which can be added to link $L_5$. This addition complements link $L_5$ by the vehicle model action took place in the delay time. In short, $L_6$ is the prediction which could be added to link $L_5$, to make it feel like undelayed feedback, link $L_7$. To do so, time-delays and vehicle model should be known. Link $L_5$ is discrete time delayed image frames which has to be transformed to feel like undelayed discrete time image frames, link $L_7$. Link $L_6$ is the forecasted position relative to the delayed position of the vehicle corresponding to the delays in the system.

It can be seen that when the vehicle model is accurate, i.e., $P' = P$, the system behaves like the one shown in figure \ref{fig:b smith}. Advantage of Smith predictor is that the delay is out of the control loop, which eliminates time delay induced instability and converts the system into just a delayed output system.

In this work, considering low vehicle teleoperation speeds, vehicle model considered is a single track kinematic model which would be discussed later in \textcolor{black}{section II-B-1-b}.


The system architecture for vehicle teleoperation is made of a control station that receives the image stream after downlink-delay ($\tau_1$) and subsequently the vehicle receives driving command after uplink-delay ($\tau_2$). Instead of showing delayed image to the human-operator a forecasted image is presented. Forecasted image tries to render the exact perspective of what vehicle would see upon reception of driving commands. For example, in case of straight driving at $4$m/s and $0.4$s of round-trip latency, the forecasting should generate the vehicle perspective at predicted position of $4$ x $0.4 = 1.6$m forward relative to the delayed image.

\subsection{Teleoperated Vehicle}
\subsubsection{Vehicle}
An electric front wheel drive vehicle, ZED One
, which is made fully actuated-by-wire system (\cite{Vignati2018}, \cite{Vignati2019}) is used for the vehicle teleoperation testing. Table \ref{tab:vehiclespec} reports, brief dimensional specification of the vehicle. \textcolor{black}{It is} equipped with camera and speed sensor that are necessary to acquire data to be transmitted to the control station.
\begin{table}[h]
	\centering
	\caption{\textcolor{black}{Vehicle Parameters}}
	\begin{tabular}{|l|l|}
		\hline
		Mass        & 668 kg  \\ \hline
		Wheelbase   & 1760 mm \\ \hline
		Track-width & 1300 mm \\ \hline
		Range       & 120 km  \\ \hline
	\end{tabular}
	\label{tab:vehiclespec}
\end{table}

\subsubsection{Camera and depth sensor}
Stereolabs ZED stereo-camera
, is used to capture stereo images and it is mounted on the vehicle roof. In normal street driving, driver has to be aware of its surroundings and their motion, to plan his course of action. \textcolor{black}{Since in majority of driving cases, the vehicle moves only in forward direction, awareness of the objects ahead is crucial role in manoeuvring vehicle}. This is why for current experiment, one available stereo camera is utilized for capturing frontward FOV.

Depth map is computed by Stereolabs SDK thanks to \emph{Nvidia Jetson AGX Xavier} installed inside the vehicle, which uses stereo images as input to generate depth-map. And later, Jetson sends depth-map and image frames to the control station.

\subsubsection{Data compression}
Since the images and depth maps are huge data to be sent through 4G network, data compression is necessary. In fact table \ref{tab:bandwidthRaw}, shows the bandwidth required to communicate uncompressed image and depth map of resolution 376x672.
\begin{table}[h]
\caption{Bandwidth required for uncompressed data}
\begin{tabular}{|l|l|c|c|c|c|}
\hline
          & Byte depth & \multicolumn{1}{l|}{Height} & \multicolumn{1}{l|}{Width} & \multicolumn{1}{l|}{\#Channels} & \multicolumn{1}{l|}{\begin{tabular}[c]{@{}l@{}}MB/s\\ @ 30 fps\end{tabular}} \\ \hline
RGB       & 1 (uint8)  & 376                         & 672                        & 3                               & \cellcolor[HTML]{FFC702} $\sim$ 22                                                                       \\ \hline
Depth-map & 4 (float)  & 376                         & 672                        & 1                               & \cellcolor[HTML]{FFC702} $\sim$ 29                                                                       \\ \hline
\multicolumn{5}{|l|}{\textbf{Total}}                                                                                         & \cellcolor[HTML]{FFC702} $\sim$ 51                                                                       \\ \hline
\end{tabular}
\label{tab:bandwidthRaw}
\end{table}
The camera resolution is chosen as a trade-off between image quality, which is necessary to perceive the details of surrounding environment, and the computational performance, which degrades as number of pixels increases.

Since the required bandwidth is much higher than the available one with 4G network, data compression is necessary (maximum theoretical upload speed in 4G is 1MB/s). JPEG compression is then utilized for RGB image compression but the same can’t be used directly for depth-map which contains floating point data. To avail benefit of JPEG compression, depth-map elements have to be converted into ‘uint8’ datatype according to the \textcolor{black}{eq. \ref{eq2}}.
\begin{align}
y=\frac{1}{A} \cdot \textcolor{black}{log}\left \{ A\left ( x-1 \right )+0.01 \right \} +C\label{eq2}
\end{align}   

\begin{figure}[ht]
\centering
\begin{subfigure}[b]{0.243\textwidth}
    \centering
    \includegraphics[width=\textwidth,cfbox=black .1pt .1pt]{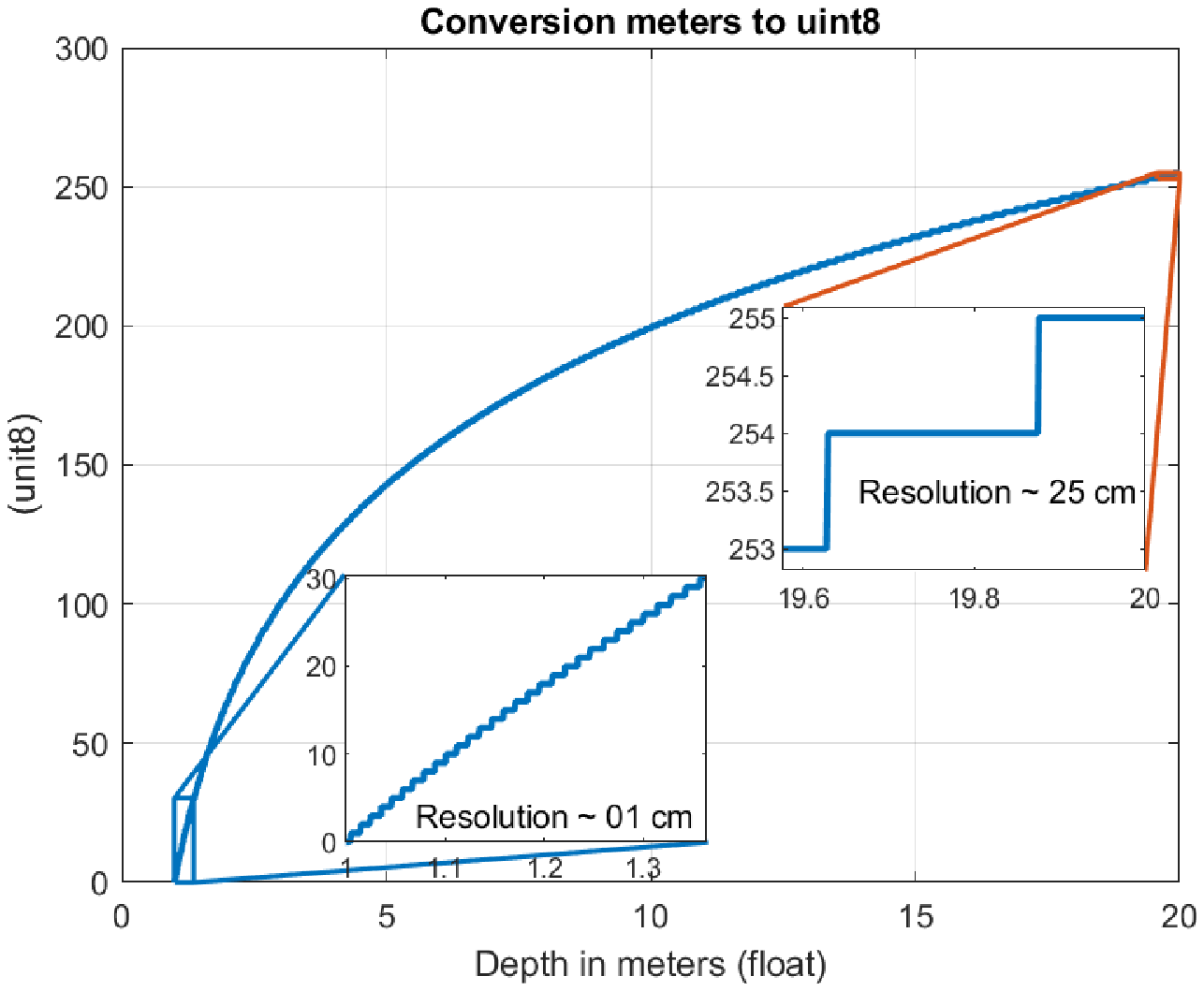}
    \caption{Depth encoding to 'uint8'}
    \label{fig:x uint8}
\end{subfigure}%
\hfill
\begin{subfigure}[b]{0.243\textwidth}
    \centering
    \includegraphics[width=\textwidth]{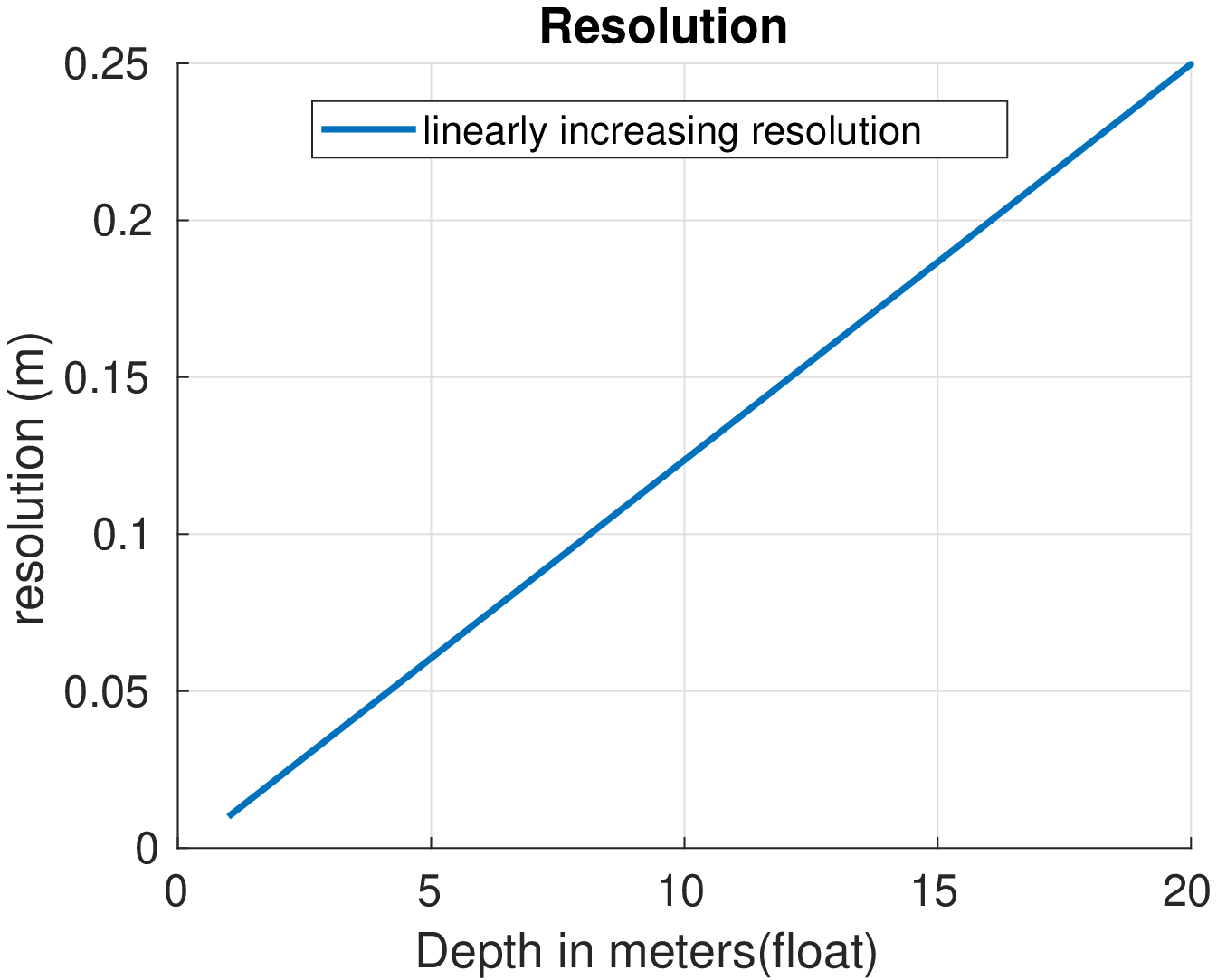}
    \caption{Linearly increasing resolution}
    \label{fig:x linear}
\end{subfigure}
\caption{Depth-map compression}
\label{fig:x depthCompression}
\end{figure}

Here, A = 0.0126194 and C = 364.92737 are obtained  to have linear increase in depth resolution with depth (figure \ref{fig:x depthCompression}). Hence, less resolution is assigned to lower depths to assure more detail for closer objects. Above relation encodes all depths (\textcolor{black}{x: }1-20 meter) in range (\textcolor{black}{y: }0-255) of ‘uint8’ data type. In control station, inverse of \textcolor{black}{eq. \ref{eq2}} is used to acquire depths in meters back again.

Exploiting this technique (eq. \ref{eq2}), JPEG compression can be utilized to further compress the depth map which now contains ‘uint8’ elements. Table \ref{tab:bandwidthCompressed} shows bandwidth required to communicate compressed image and depth map of resolution 376x672. Data compression makes it feasible to use 4G network for wireless data communication.
\begin{table}[H]
\caption{Bandwidth required for compressed data}
\begin{tabular}{|l|l|c|c|c|c|}
\hline
          & Byte depth & \multicolumn{1}{l|}{Height} & \multicolumn{1}{l|}{Width} & \multicolumn{1}{l|}{\#Channels} & \multicolumn{1}{l|}{\begin{tabular}[c]{@{}l@{}}MB/s\\ @ 30 fps\end{tabular}} \\ \hline
RGB       & 1 (uint8)  & 376                         & 672                        & 3                               & \cellcolor[HTML]{34FF34}$\sim$ 0.70                                                                     \\ \hline
Depth-map & 1 (uint8)  & 376                         & 672                        & 1                               & \cellcolor[HTML]{34FF34}$\sim$ 0.40                                                                     \\ \hline
\multicolumn{5}{|l|}{\textbf{Total}}                                                                                         & \cellcolor[HTML]{34FF34}$\sim$ 1.10                                                                     \\ \hline
\end{tabular}
\label{tab:bandwidthCompressed}
\end{table}

\subsection{Control-station}
Control-station consists of a PC powered by intel i7 processor, 24" display, Logitech G920 steering wheels, and pedals as shown in figure \ref{fig:x controlStation}. Due to the constrain of live streaming, time available to process each image frame is $\sim$33ms (for 30fps). For faster processing, CPU parallel computing is utilized to perform the perspective transformation. Due to which processing time for each frame is $\sim$15ms. The human-operator sees the transformed image and actuates steering wheel. Human operator can also see current speed, estimated path (based on current steering angle) and observed network latency for situational awareness.

Control-station tasks are predicting forecasted position corresponding to delay in the control loop, performing perspective projection to generate the new perspective and capturing operator commands to transmit it back to the vehicle.

\subsubsection{Position prediction}
\begin{figure}[ht]
\centering
\includegraphics[width=0.4\textwidth,cfbox=black .1pt .1pt]{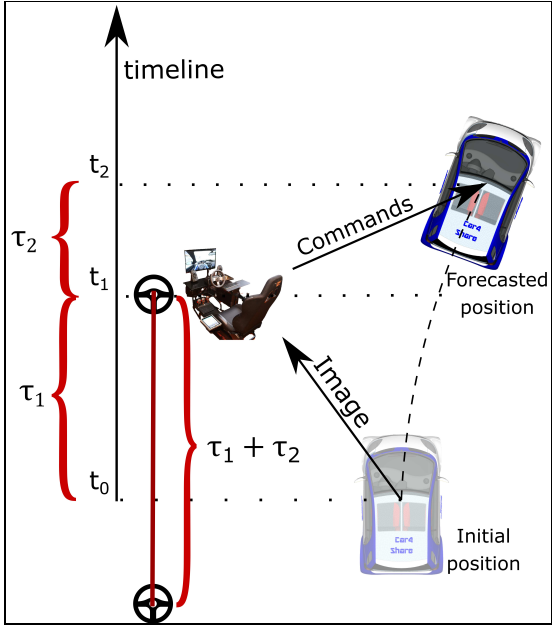}
\caption{Concept of forecasting position}
\label{fig:x predictionConcept}
\end{figure}

Forecasted position is the position, where vehicle would receive driving commands back, respective to the input image frame it sent to the control-station (see figure \ref{fig:x predictionConcept}). Control-station receives image frame, depth frame, vehicle speed, and acceleration. Forecasting time-window depends on time-delay present in data communication. Taking control-station as reference point, downlink-delay \textcolor{black}{(eq. \ref{eq4})} is considered as the delay associated with image frame. Uplink-delay \textcolor{black}{(eq. \ref{eq5})} is considered as the delay associated with driving commands. \textcolor{black}{It consists of camera capture delay, depth-map computation delay, data communication delay, and PP computation time. These contributions can be lumped, because time point of image capture is used in image timestamp. Uplink-delay consists of network uplink delay, and vehicle actuation control unit delay. It is unknown during image transformation at control-station. }Stochastic approach is considered to estimate the uplink-delay.

\renewcommand{\theenumi}{\alph{enumi}}
 \begin{enumerate}
   \item Uplink-delay estimation: \textcolor{black}{Since actuation delay($\sim20ms$) is constant, it is combined directly with $\tau_1$. Here onward only network part of uplink delay is considered as net uplink delay ($\tau_2$), which can only be observed by the vehicle.}
Making reference to figure \ref{fig:x latencyCheck}, the uplink-delay measured during a vehicle teleoperation test is reported as occurrence histogram. Analyzing the collected data of 9000 data-points of uplink-delay, it was found that \emph{Generalized extreme value (GEV) distribution} fits accurately on uplink time-delay data. Also, Y. Zheng \emph{et al}. \cite{Zheng2020} found GEV distribution apt for representing distribution of time-delays in  mobile communication. The probability distribution of generalized extreme value distribution is defined by  \textcolor{black}{eq. \ref{eq gev}}.
\begin{align}
f(\textcolor{black}{\tau_2})_{\xi\neq0, \mu, \sigma} =  \frac{1}{\sigma} \left(1+\xi \frac{\textcolor{black}{\tau_2}-\mu}{\sigma} \right)^{-\frac{1}{\xi}-1}e^{-\left(1+\xi \frac{\textcolor{black}{\tau_2}-\mu}{\sigma} \right)^{-\frac{1}{\xi}} }\label{eq gev}
\end{align}   

For $1+\xi\left(\frac{\textcolor{black}{\tau_2}-\mu}{\sigma}\right)>0$, $\xi$ the shape parameter, $\mu$ the location parameter and $\sigma>0$ the scale parameter. The shape parameter $\xi$, is found positive. Which means, the distribution has a lower bound $\mu-\frac{\sigma}{\xi}$ and heavy tail, based on the extreme value theory. \textcolor{black}{Every second, control-station fits GEV distribution (using MATLAB $gevfit$) on latest uplink-delay data, computes its $95^{th}$ and $99.9^{th}$ percentile (using MATLAB $gevinv$) and send it to control-station}.
\begin{figure}[ht]
\centering
\includegraphics[width=0.4\textwidth]{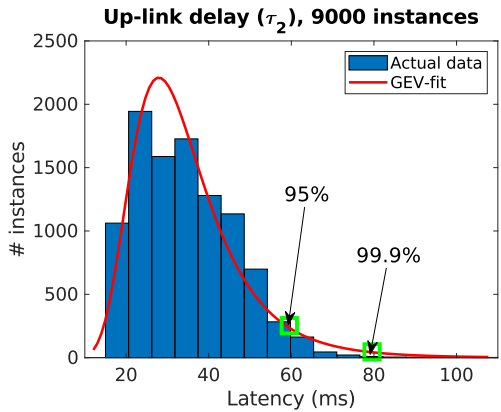}
\caption{GEV distribution fit for uplink-delay data}
\label{fig:x latencyCheck}
\end{figure}

 \textcolor{black}{J. C. Lane \emph{et al}. \cite{Corde:2002} found that a short variable lag could be more determinantal than a longer fixed one in a human-in-the-loop system. To eliminate the variability factor in the uplink-delay, hold and apply strategy is used.} Consequently in the control station, for forecasting vehicle position, instead of considering most probable delay value of the uplink-delay distribution, $95^{th}$ percentile of distribution is considered. \textcolor{black}{For instance, if the most probable delay is 30 ms and delay corresponds to $95^{th}$ percentile is 60 ms. During forecasting position, uplink-delay considered is 60 ms. The human operator would maneuver according the the forecasted vehicle position and the generated driving commands would correspond to 60 ms uplink-delay. Considering higher percentile instead of most probable value for uplink delay, ensures prior reception of driving commands by vehicle compare to its desired actuation time ( in $95	\%$ instances).} Besides containing the driving commands, the uplink msg consists of three time-info's. First time-info corresponds to the control-station timestamp when the driving command was generated by the human operator. Second and third time-info's consists of $95^{th}$ and $99.9^{th}$ percentile of delay distribution respectively. The vehicle is programmed to direct the commands to its actuators at the time \lq control-station timestamp + $95^{th}$ percentile of delay distribution\rq. \textcolor{black}{Hold and apply strategy narrows the effective variability of uplink-delay.} Purpose of computing $99.9^{th}$ percentile is to inform the vehicle to activate safety actions, if vehicle doesn't receive any command even waiting till the time corresponds to $99.9^{th}$ percentile. In case of weak network condition, distribution fit may get widen. Which may result much higher value for $99.9^{th}$ percentile. A max limit of 200 ms is set for $99.9^{th}$ percentile. If vehicle doesn't receive any command till max 200 ms from the time-stamp of last command, emergency stop manoeuvre would be activated by the vehicle. Figure \ref{fig:x uplinkTrend} shows the trend of $95^{th}$ percentile and $99.9^{th}$ percentile of uplink-delay over a time window of 90s. Here the distribution fit is performed every second over the 50 uplink-delays observed in each second, as the control-station transmits command at $50$ hz. It can be seen that variability of the $95^{th}$ percentile (used in hold and apply strategy) has been reduced to $\sim10$ms. 
 Since the $99.9^{th}$ percentile is not directly involve in hold and apply strategy, its variability has less impact on the whole operation.
 \begin{figure}[ht]
\centering
\includegraphics[width=0.4\textwidth]{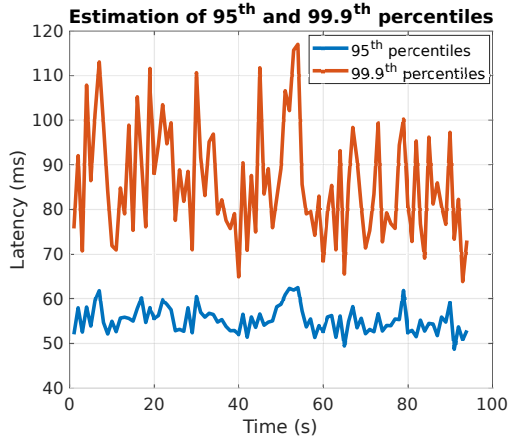}
\caption{Estimation of $95^{th}$ and $99.9^{th}$ percentiles of uplink-delays over a time window of 90 s}
\label{fig:x uplinkTrend}
\end{figure}

\item Trajectory integration to predict vehicle position: Single track kinematic model is used to integrate vehicle trajectory. \textcolor{black}{Feng. Gao \emph{et al}. \cite{Gao2021} found that at low lateral acceleration ($<2m/s^2$) and at non-small steering angles, kinematic vehicle model is comparable to non-linear coupled vehicle dynamics model. With trajectory integration, rear-axle center pose change is computed. Then, camera (fig \ref{fig:x depiction2}) pose change is computed with the help of rear-axle center pose change.}

Time-delay in wireless communication is variable. To make Smith predictor approach valid for variable time delays, instantaneous downlink delay value is considered instead of mean value of downlink delay. To account instantaneous value of downlink delay, a full integration of the vehicle predictor model from time \textcolor{black}{ $t_1-(\tau_1+\tau_2)$ to $t_1$} is performed for every image frame received at the control station, see figure \ref{fig:x predictionConcept}.
\begin{align}
&Downlink\ delay,~\tau_1&=  t_1-t_0\label{eq4} \\
&Uplink\ delay,~\tau_2&=  t_2-t_1\label{eq5}\\
&Total\ delay,~\tau&=  \tau_1 + \tau_2\label{eq6}   
\end{align}
\textcolor{black}{At time instant $t_1$, when control-station receives an image frame, steering time-history of control-station (link $L_2$ in figure \ref{fig:a smith}) is known.} Euler method, Runge-Kutta method or Exact trajectory integration method can be used for forward kinematic integration. Here, exact trajectory integration method is used with steering time-history of time-window equal to $\tau$. Output of trajectory integration is the pose change of the camera relative to its input pose as shown in figure \ref{fig:x depiction2}. Intermediate output of trajectory integration is the pose change of rear axle center with respect to its input pose (Input pose of rear axle is always lies on the origin as given in eq \ref{eq7}). Image coordinate system is used to maintain relation between vehicle movement and image captured by camera.
Considering initial conditions:
\textcolor{black}{
\begin{flalign}
&X_0=Z_0=\Psi_0=0&\label{eq7}\\
&V_0=Speed\;(Received\;at\;time = t_1-\tau\;)&\label{eq8}\\
&a\;\;=Acceleration\;(Received\;at\;time = t_1-\tau\;)&\label{eq9}
\end{flalign}
}
\textcolor{black}{Where, $\tau$ is given by eq \ref{eq6}}. To predict the curved path of the vehicle, the time history of the steering is needed.

Steering time-history is:
$\begin{bmatrix}
\delta_0 &dt_0\\
\delta_1 &dt_1\\
\cdot &\cdot\\
\delta_{n-1} &dt_{n-1}\\
\end{bmatrix}$\\
Here, first column contains the steer angles at front wheel and second column contains respective time intervals for which the steer is effective. Due to constant sampling \textcolor{black}{time of $0.020s$}, most of the time intervals are constant, except $dt_0$ and $dt_{n-1}$. This is due to absence of synchronicity between steering angle sampling and image streaming frame rate.

\textcolor{black}{Equation \ref{eq10}} calculates the velocity of rear axle center at each integration step.
\begin{flalign}
&V_i=V_0 + a\sum_{n=0}^{i-1}dt_n&\label{eq10}
\end{flalign}

Given the steer, instantaneous radius of curvature ($R$) and angular velocity ($\omega$) is given by eq \ref{eq11}-\ref{eq12}, where $L$ is the vehicle wheelbase. 

\begin{flalign}
&R_i = \dfrac{L}{tan\,\delta_i}&\label{eq11}
\\
& \omega_i  = \dfrac{V_i}{R_i}&\label{eq12}
\end{flalign}   

Equation \ref{eq13}-\ref{eq15} gives the pose of rear axle center after $i^{th}$ integrating step, according to reference frame shown in figure \ref{fig:x depiction2}, where origin lies at the center of rear axle.

\begin{flalign}
&   \Psi  _{i+1}  = \Psi _{i} + \omega_i \cdot dt_i&\label{eq13}
\\
&X_{i+1}=
\begin{cases}
    X_{i}+(V_i\ dt_i)\cdot \sin \Psi _{i}& \text{if }  \omega  =  0\\
    X_{i}-R(\cos \Psi  _{i+1} - \cos \Psi  _{i})              & \text{otherwise}
\end{cases}&\label{eq14}
\\
&Z_{i+1}=
\begin{cases}
    Z_{i}+(V_i\ dt_i)\cdot \cos \Psi _{i}&\;\; \text{if }  \omega  =  0\\
    Z_{i}+R(\sin \Psi  _{i+1} - \sin \Psi  _{i})              &\;\; \text{otherwise}
\end{cases}&\label{eq15}
\end{flalign}   

After performing integration over the steering time-history window, predicted pose ($X_n, Z_n, \Psi _n$) of rear-axle center is obtained.

Camera pose change \textcolor{black}{($\Delta Z_{cam}, \Delta X_{cam}, \Delta \Psi_{cam}$)}, i.e., relative pose of predicted camera position with respect to input camera position \textcolor{black}{(figure \ref{fig:x depiction2})} is obtained through \textcolor{black}{eq. \ref{eq16}-\ref{eq18}}.
\begin{align}
\Delta Z_{cam}&=Z_n-C \cdot (1-\cos\Psi_n)\label{eq16}\\
\Delta X_{cam}&=X_n+C \cdot \sin\Psi_n\label{eq17}\\
\Delta \Psi_{cam}&=\Psi_n\label{eq18}
\end{align}   
Here, $C$, is horizontal distance between camera and the rear axle center.
Camera pose change is required because the link $L_5$ in figure (\ref{fig:a smith}) is the image captured with camera mounted on the vehicle. Camera pose change is the link $L_6$ in figure (\ref{fig:a smith}).

\end{enumerate}

\subsubsection{Perspective Projection}
\begin{figure}[h]
\centering
\includegraphics[width=0.4\textwidth]{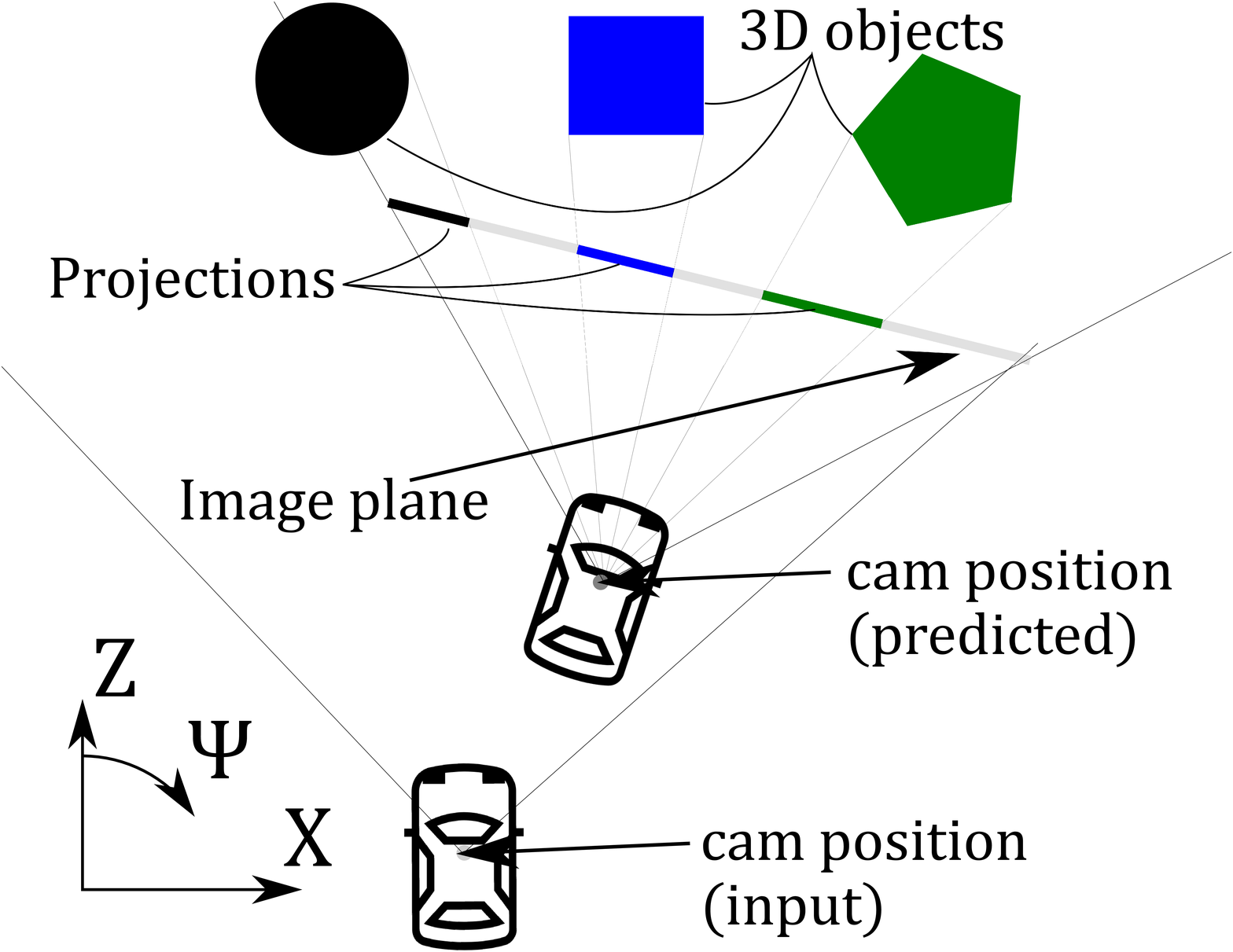}
\caption{A simple 2D representation of Perspective Projection}
\label{fig:x depiction2}
\end{figure}

Perspective projection refers to image transformation to obtain the new camera perspective after the camera \textcolor{black}{has} traversed. It corresponds to the 3D projection of the world on the image plane of the camera at the predicted position and its topology perceived by the camera at its predicted position as depicted in figure \ref{fig:x depiction2}. Unlike typical zooming, it resizes each object ahead considering its distance from the camera. It also captures the effect of vehicle yaw motion on the new image formed, as shown in figure \ref{fig:x depiction2}. Through this technique, human-operator sees a forecasted video stream which tries to replicate future perspective of vehicle FOV. It forecasts by accounting measured time-delay, vehicle speed, and steering time-history.

Object's new perspective depends upon its relative coordinates from the camera. Nearer objects expand more than farther objects. While making a turn, some objects disappear and new objects appear. Object disappearance can be simulated well but appearance of new object can not be simulated. Inpainting technique is used to fill those pixels which correspond to introduction of new objects. Algorithm proposed by Telea \cite{Telea} is used for inpainting, which considers the color information of available neighbouring pixels. To compute relative position of objects, depth-map is used as input for this transformation. Depth-map, is an image channel that contains information relating to the distance of the surfaces of scene objects from the viewpoint. The output of this projection is the link $L_7$ in figure (\ref{fig:a smith}). Process flow chart of perspective projection implementation is shown in figure (\ref{fig:x PITblocks}). \textcolor{black}{Errors correspond to each block is discussed in Appendix A, where the prime contributor is the \textit{Point-cloud transformation} block.}

\begin{figure}[h]
\centering
\includegraphics[width=0.49\textwidth]{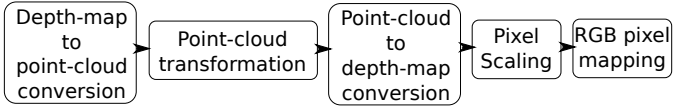}
\caption{Process flow chart of Perspective Projection}
\label{fig:x PITblocks}
\end{figure}

\renewcommand{\theenumi}{\alph{enumi}}
 \begin{enumerate}
   \item Inputs: Image, depth map, camera FOV angles (horizontal and vertical both), camera displacements ( $\Delta Z_{cam}$, $\Delta X_{cam}$ and $\Delta \Psi_{cam}$ ) are the inputs to the perspective transformation algorithm. Input image is considered perfectly rectified from lens distortion. Depth map has the same resolution as of the input image and every element in depth map corresponds exactly to the respective pixel in input RGB image. Ideally, point-cloud shall be used as an input to Perspective projection, but as point-cloud is high bandwidth consuming data, depth map is used instead of point-cloud.

\item Conversion of Depth-map into Point-cloud: After receiving depth map of old perspective, control-station converts it into point-cloud.
Two approaches can be used for this conversion. First, using camera focal lengths in x and y. Second, using the horizontal and vertical FOV of the depth-map image. Since, image and depth map are free of lens distortion, FOV approach is used. Equations (\ref{eq19}-\ref{eq20}) are used to convert depth map into point-cloud. $W$ and $H$ are width and height of the image in pixels.

Say, a point in depth-map is at $x_d$-col and $y_d$-row (\textcolor{black}{ranges are 1 to W and 1 to H respectively}) has a depth value of $\hat z_p$ meters. First, shift the depth-map origin from top left corner to the center of the depth-map. This is done since image reference frame origin is in the top-left corner while the optical axis of the camera view passes through the middle of the image plane.
\begin{align}
\hat x_d = x_d-(W/2+0.5)\nonumber \\
\hat y_d = y_d-(H/2+0.5)\label{eq19}
\end{align}   
\textcolor{black}{Equation \ref{eq20} computes the Cartesian coordinates ($\hat x_p, \hat y_p$) of the pixel in meters.}
\begin{align}
\hat x_p = \hat z_p \left [ \hat x_d\left \{ \frac{tan(\frac{fovH}{2})}{W/2} \right \} \right ]\nonumber \\
\hat y_p = \hat z_p \left [ \hat y_d\left \{ \frac{tan(\frac{fovV}{2})}{H/2} \right \} \right ]\label{eq20}
\end{align}   
\\
Where, $fovH$ and $fovV$ are horizontal and vertical FOV of the depth-map image. Needless to say that z coordinate of the point is already known, as it is given in that particular depth map point. $(\hat x_p, \hat y_p, \hat z_p)$ is the point in point-cloud corresponds to $(x_d, y_d)$ pixel in depth-map.

\item Point-cloud transformation: The above point-cloud has origin on input position of the vehicle (refer \textcolor{black}{figure \ref{fig:x depiction2}}, cam position input). It has to be transformed into the new reference frame corresponds to predicted position. Thus, a rotation and translation to transform the point-cloud from cam position input to cam position predicted (refer \textcolor{black}{figure \ref{fig:x depiction2}}) are required.

Rotation is performed according to the following rotation matrix \textcolor{black}{(eq \ref{eq21})},
\begin{flalign}
&R = 
\begin{bmatrix}
cos(\Delta \Psi_{cam}) &0 &sin(\Delta \Psi_{cam})\\
0 &1 &0\\
-sin(\Delta \Psi_{cam}) &0 &cos(\Delta \Psi_{cam})
\end{bmatrix}&\label{eq21}
\end{flalign}   

Above matrix is based on coordinate system indicated in Figure \ref{fig:x depiction2}. It means that a rotation is positive while turning right.

\begin{flalign}&\text{The translation vector, }D =
\begin{bmatrix}
\Delta X_{cam}\\
0\\
\Delta Z_{cam}
\end{bmatrix}&\label{eq22}
\end{flalign}   
\\
Previous two operations can be combined in a homogeneous transformation matrix \textcolor{black}{(eq \ref{eq23})}, 
\begin{flalign}&T =
\begin{bmatrix}
R^T &-R^T\cdot D\\
0  &1
\end{bmatrix}&\label{eq23}
\end{flalign}   

In our case, camera is mounted pointing slightly downward, for better utilization of vertical FOV as shown in figure \ref{fig:x pitchAng}. The transformation matrix T has then to be modified as \textcolor{black}{per eq \ref{eq24} and eq \ref{eq25}}, accounting for the additional rotation about x axis.
\begin{flalign}&\text{The rotation matrix, }R_2 = 
\begin{bmatrix}
1 &0 &0\\
0 &cos(\theta) &-sin(\theta)\\
0 &sin(\theta) &cos(\theta)\\
\end{bmatrix}&\label{eq24}
\end{flalign}   
\\
New transformation matrix, 
\begin{align}
T_2 = 
\begin{bmatrix}
R_2 &0\\
0 &1
\end{bmatrix}
\cdot
T
\cdot
\begin{bmatrix}
R_2^T &0\\
0 &1
\end{bmatrix}\label{eq25}
\end{align}   
\\
\begin{figure}[h]
    \centering
    \includegraphics[width=0.4\textwidth]{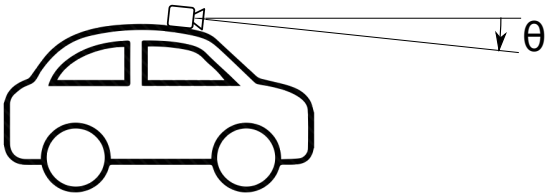}
    \caption{Camera downward inclination}
    \label{fig:x pitchAng}
\end{figure}
\\
New point coordinates ($\hat x_{pNew},~\hat y_{pNew},~\hat z_{pNew}$) corresponding to the old coordinates ($\hat x_{p},~\hat y_{p},~\hat z_{p}$) are given by \textcolor{black}{eq \ref{eq26}}:
\begin{align} 
\begin{bmatrix}
\hat x_{pNew}\\
\hat y_{pNew}\\
\hat z_{pNew}\\
1
\end{bmatrix}
=
T_2
\cdot
\begin{bmatrix}
\hat x_{p}\\
\hat y_{p}\\
\hat z_{p}\\
1
\end{bmatrix}\label{eq26}
\end{align}   
New point coordinates ($\hat x_{pNew},~\hat y_{pNew},~\hat z_{pNew}$) is in reference frame of camera of predicted position of the vehicle (refer \textcolor{black}{figure \ref{fig:x depiction2}}).

\item Remapping of point-cloud to depth-map: \textcolor{black}{Once the new Cartesian coordinates ($\hat x_{pNew},~\hat y_{pNew},~\hat z_{pNew}$) of the pixel is computed, it is again converted back to depth-map, to get a new depth-map with eqs \ref{eq27}-\ref{eq28}.}

\begin{align}
\hat x_{dNew} = \frac{\hat x_{pNew}}{\hat z_{pNew}} \left \{ \frac{W/2}{tan(\frac{fovH}{2})} \right \}\label{eq27}
\end{align}   

\begin{align}
\hat y_{dNew} = \frac{\hat y_{pNew}}{\hat z_{pNew}} \left \{ \frac{H/2}{tan(\frac{fovV}{2})} \right \})\label{eq28}
\end{align}   
Above obtained ($\hat x_{dNew}, \hat y_{dNew}$) coordinates are new pixel location corresponds to the old pixel location ($\hat x_{d}, \hat y_{d}$). Right now it is centered with image center. Re-centering back to the top-left corner is performed as \textcolor{black}{given by eq \ref{eq29}}.
\begin{align}
x_{dNew} = \hat x_{dNew}+(W/2+0.5)\nonumber \\
y_{dNew} = \hat y_{dNew}+(H/2+0.5)\label{eq29}
\end{align}   
This is necessary because in image pixel coordinate system, origin is at the top-left corner. Although in this example, only one pixel is discussed but the same has to be done for all the pixels. The output of this step is a map which contains information about the new location ($x_{dNew},~y_{dNew}$) of the old pixel ($x_{d}, y_{d}$) in the the image.
\item Pixel scaling:
Pixel scaling performs re-scaling the objects in image according to their new distance from the camera, i.e., nearer objects would scale-up more compare to farther objects. This scaling is performed at the pixel level. Scale factor for each pixel depends on old and new depth of that pixel. Scale factor for pixel ($x_{dNew},~y_{dNew}$) is \textcolor{black}{given by eq \ref{eq30}}.
\begin{align}
S = \frac{\hat z_p}{\hat z_{pNew}}\label{eq30}
\end{align}   
As scale factor is different for each pixel, span (or spread) of
each pixel in new image would be different.\\
Span of old pixel ($x_{d}, y_{d}$) in new image over the rows is \textcolor{black}{given by eq \ref{eq31}}.
\begingroup\makeatletter\def\f@size{8}\check@mathfonts
\def\maketag@@@#1{\hbox{\m@th\large\normalfont#1}}%
\begin{align}
iRows = \left [ floor\left (  y_{dNew}-\frac{S-1}{2} \right ) : ceil\left (  y_{dNew}+\frac{S-1}{2} \right ) \right ]\label{eq31}
\end{align}   
\endgroup
Span of old pixel ($x_{d}, y_{d}$) in new image over the columns is \textcolor{black}{given by eq \ref{eq32}}.
\begingroup\makeatletter\def\f@size{8}\check@mathfonts
\def\maketag@@@#1{\hbox{\m@th\large\normalfont#1}}%
\begin{align}
iCols = \left [ floor\left (  x_{dNew}-\frac{S-1}{2} \right ) : ceil\left (  x_{dNew}+\frac{S-1}{2} \right ) \right ]\label{eq32}
\end{align}   
\endgroup
\\
\item Mapping RGB values for each pixel: Being known the old location and new location of each pixel, RGB information is carried from old image to the new image. Psuedo code for carrying RGB information for one old pixel is \textcolor{black}{given by eq \ref{eq33}}.
\begin{equation}\label{eq33}
\resizebox{0.9\hsize}{!}{$newImage(iRows, iCols, 1:3) = oldImage(y_d , x_d , 1:3)$%
}%
\end{equation}
Here $(x_d, y_d)$ are the inputs used in $RHS$ in eq \ref{eq19}. Mapping of RGB pixels is responsible to place objects in previous FOV to new FOV, i.e., to place RGB pixels $(x_d, y_d)$ from figure \ref{fig:x input} to $( x_{dNew},  y_{dNew})$ in figure \ref{fig:x output}.
To get the full new image, new locations of all the old pixels are calculated. Obviously, the pixels that fall out of frame need to be discarded. Important thing to keep in mind during this iteration is, the order in which pixels need to be processed. As can be imagined, when camera moves forward, nearer objects scale-up and try to blanket farther objects. Therefore, pixels which correspond to farther depths need to be processed before than pixels which correspond to closer depths.
\\
\item Performance check: To verify the performance of Perspective projection algorithm, an input RGB image (figure \ref{fig:x input}) and corresponding depth-map is passed into it. Predicted position of vehicle is considered at $\Delta X=0.5m,~\Delta Z=2.1m,~\Delta \Psi=15^{\circ}$.
\\
\begin{figure}[h]
\centering
\begin{subfigure}[b]{0.24\textwidth}
    \centering
    \includegraphics[width=\textwidth]{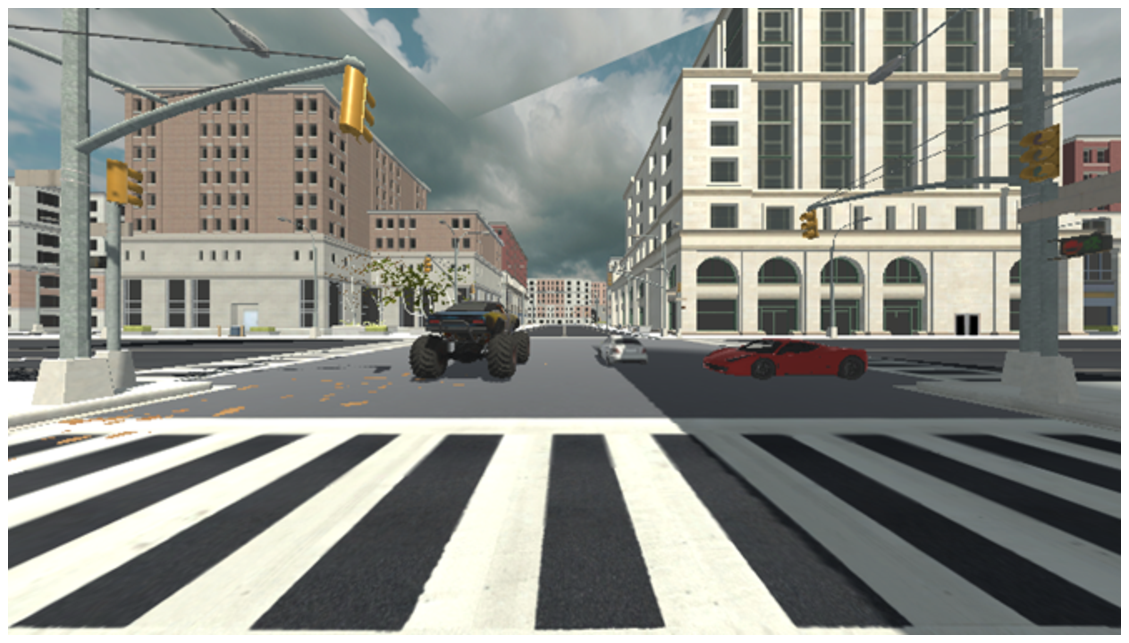}
    \caption{Input image}
    \label{fig:x input}
\end{subfigure}\\[1ex]
\begin{subfigure}[b]{0.24\textwidth}
    \centering
    \includegraphics[width=\textwidth]{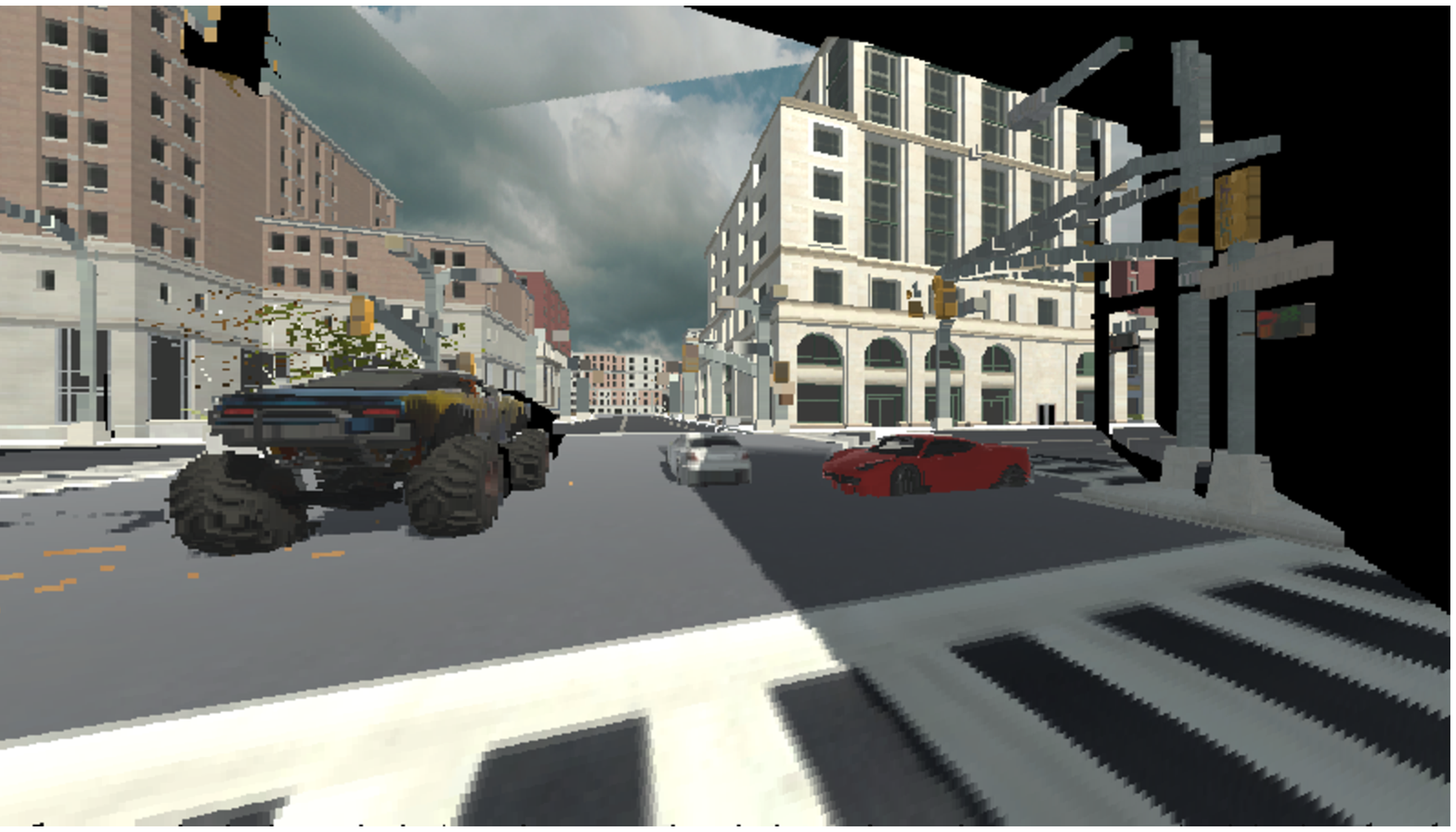}
    \caption{Output image}
    \label{fig:x output}
\end{subfigure}%
\hfill
\begin{subfigure}[b]{0.24\textwidth}
    \centering
    \includegraphics[width=\textwidth]{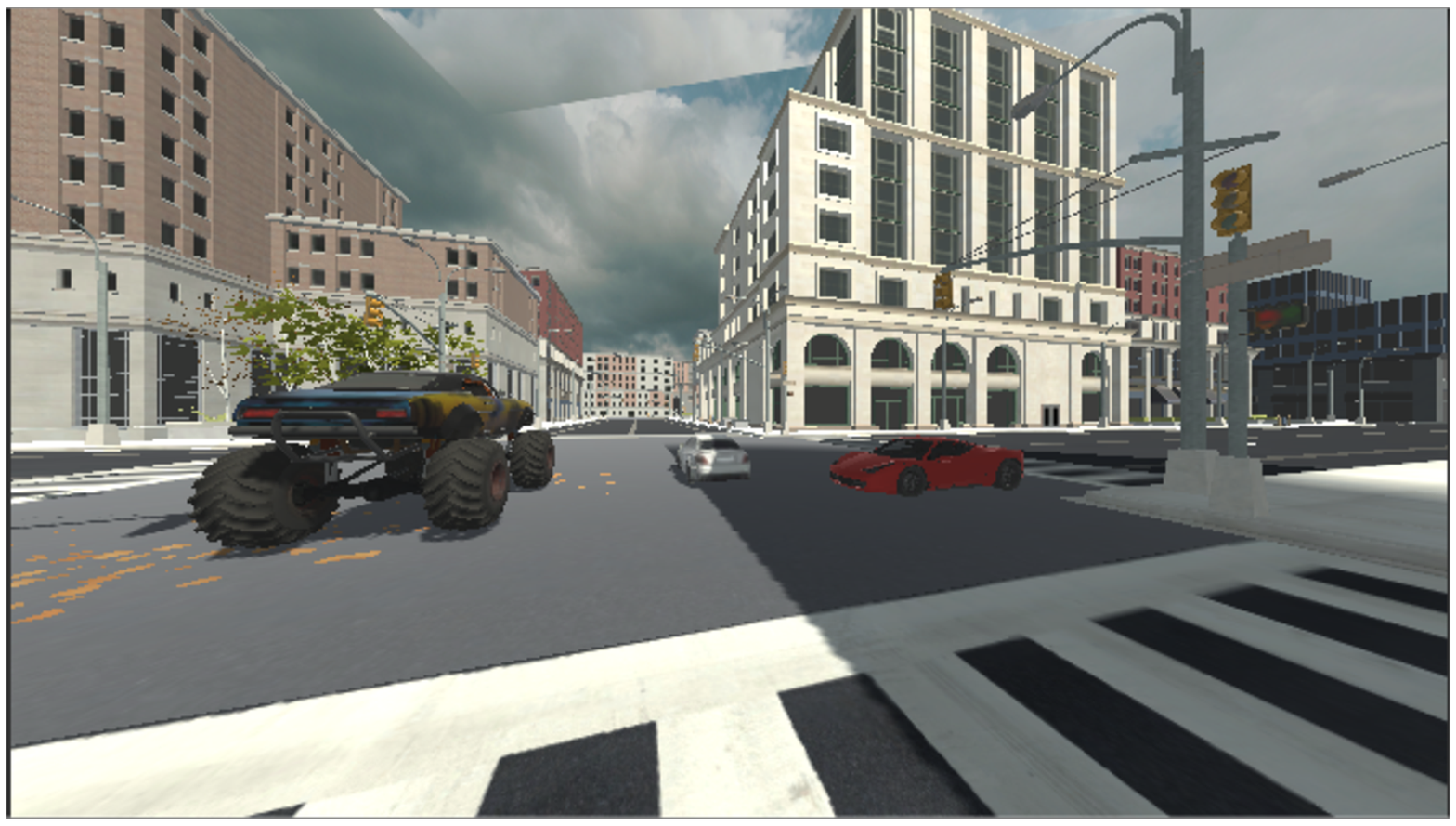}
    \caption{Ground truth}
    \label{fig:x groundTruth}
\end{subfigure}
\caption{Performance check of PP}
\label{fig:x performance}
\end{figure}

Figure \ref{fig:x output}, shows output obtained after perspective projection of input image figure \ref{fig:x input}. Output image can be compared with ground truth figure \ref{fig:x groundTruth}. Ground truth corresponds to the perspective of the vehicle, when vehicle reached at the predicted position. It can be noticed that sizes of farther objects didn't change significantly, but sizes of closer objects changed significantly (e.g. the size off-road vehicle in image has scaled-up significantly). Objects in the left part of the image is automatically discarded in the frame because they are no longer belong to the new FOV. Right part of the image (figure \ref{fig:x output}) is \textit{null} because new objects can not be predicted by this approach (\textcolor{black}{black pixels in }figure \ref{fig:x output}). Obtained output can be compared with effective view shown in figure \ref{fig:x groundTruth} after vehicle motion. In general good matching is observed.

\end{enumerate}

\subsection{Network communication}
To teleoperate the vehicle, a LAN is generated connecting the vehicle with the control station. This is made thanks to Robot Operating System (ROS) network which is set up with university VPN over 4G communication using ‘Netgear MR1100’ modem. The modem is LTE Category 16 which is compliant with 3GPP Release 12. So, it could avail LTE Advanced technology also, which is a major enhancement of the LTE standard.

\subsection{GPS}
To track vehicle trajectory, GPS \emph{AgGPS-332} receiver is mounted on vehicle. GPS controller is constantly performing RTK-correction to ensure \emph{centimeter} level accuracy, to have good trajectory analysis. To compare the trajectories traversed in the following section GPS tracked Geo-coordinates are used.
\section{Results}
\begin{figure}[ht]
\centering
\includegraphics[width=0.5\columnwidth]{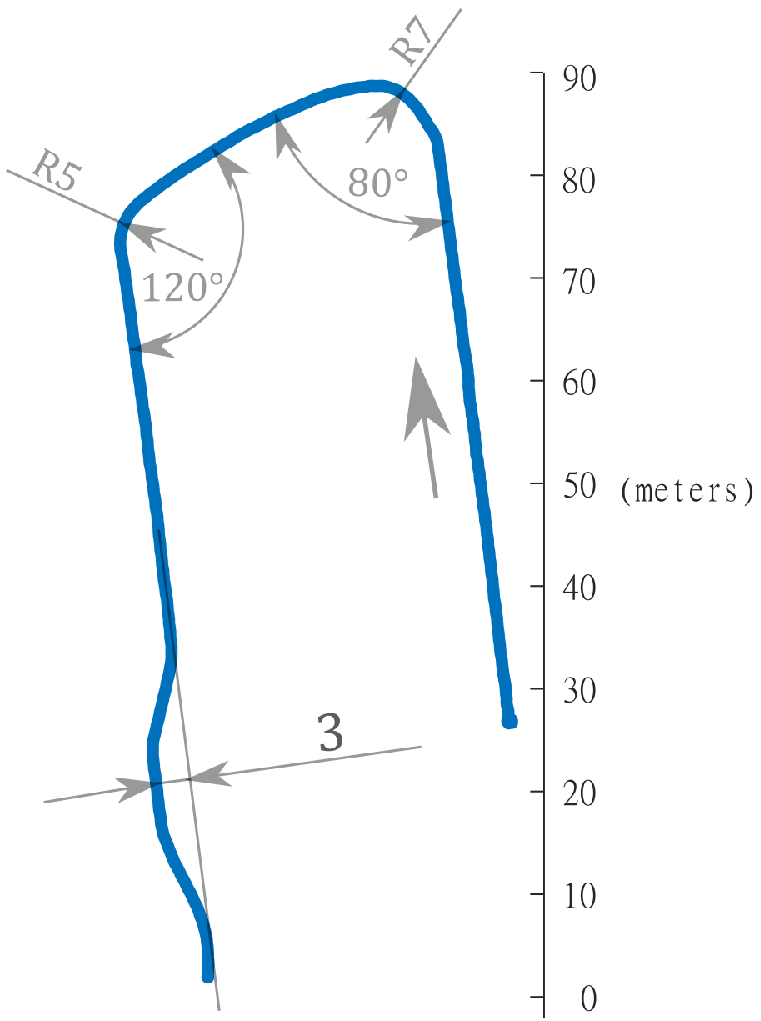}
\caption{Track for experiment}
\label{fig:x manoeuvre}
\end{figure}

To assess the usefulness of perspective projection technique, some on-street edge case manoeuvres are performed in controlled environment. Manoeuvres considered for this assessment are 7 meter radius \ang{80} left-angled turn (R7-\ang{80}), 5 meter radius \ang{120} left-angled turn (R5-\ang{120}) and double lane change (as shown in figure \ref{fig:x manoeuvre}). Test track is an asphalt rolled road with center-line marked throughout with reflective wide tape. Marked center-line hints the driver about the current deviation while driving and also the rms value of deviation is being used to assess the performance of Perspective projection in teleoperation.

In this experiment two people are involved. One person is inside the vehicle acting as a safety driver and other is in the control-station for teleoperation laps. Experiment is performed in 3 different modes:
\begin{enumerate}
\item Driver in vehicle: The driver inside the vehicle drives on the track at constant speed of $10$ km/h to capture GPS coordinates of the track while normal vehicle driving.
\item Teleoperation with perspective projection: In control-station, after performing 20 minutes of drive training on above manoeuvres with perspective projection algorithm, the human operator perform two teleoperation laps with the algorithm at constant speed of $10$ km/h.
\item Teleoperation without perspective projection: In control-station, after performing 20 minutes of drive training on above manoeuvres without perspective projection algorithm, the human operator perform two teleoperation laps without the algorithm at constant speed of $10$ km/h.
\end{enumerate}

Short training time (of only 20 minutes) was considered intentionally  to assess the performance benefit of perspective projection algorithm in spontaneous scenarios. Considering longer training times may affect the assessment due the fact that human mind tries to correct, learn and memorize repetitive control actions.

To evaluate the performance benefits, trajectories are monitored with the help of two GPS antennas mounted at the front and rear of the vehicle roof top. With the help of two GPS antennas vehicle heading also can be precisely tracked. While driving at low speeds in this experiment, general tendency of driver is to coincide forward-most point of the car with the center line of road. Forward-most point of the car is found at $0.8$ m ahead from the front axle. Geographic coordinates of this forward-most point is estimated through linear extrapolation of the GPS coordinates from the two GPS antenna mounted on the vehicle roof top. RMS deviation - To assess performance improvement, deviation is measured between this foremost point of the vehicle and closest point on center-line of the test track. Tests are performed at only one speed of 10 km/h for all manoeuvres. Constant speed is maintained by a cruise control system installed inside the vehicle. Whereas, steering commands are generated by the human inputs, transmitted from the control-station. \textcolor{black}{During all teleoperation laps, the safety driver didn't intervene}.
\\
RMS deviation is computed \textcolor{black}{as given by eq \ref{eq34}}.
\begin{align}
&RMSE,\, \varepsilon = \sqrt{\dfrac{\sum deviation^2}{n_{GPS}}}\label{eq34}
\end{align}   
where,
\\
$deviation$ - is the \textcolor{black}{minimum} distance between forward-most point and the track
\\
$n_{GPS}$ - is the number of GPS readings observed in the specific manoeuvre

\begin{figure}[ht]
\centering
\includegraphics[width=1\columnwidth,cfbox=black .1pt .1pt]{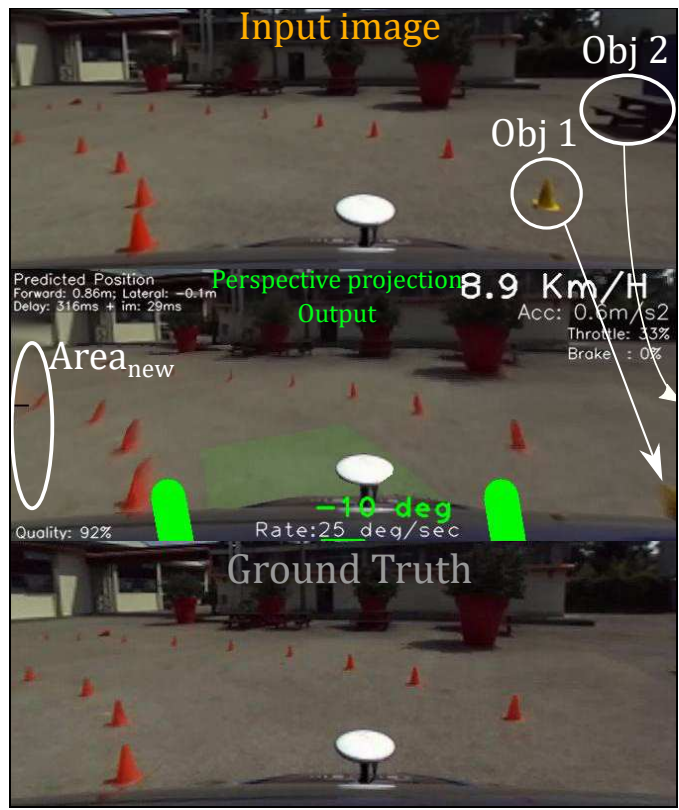}
\caption{Input and output image of Perspective projection node}
\label{fig:x inputOutputPIT}
\end{figure}

\textcolor{black}{Figure \ref{fig:x inputOutputPIT} shows the input, output image of the Perspective projection node and ground truth (a snap from \textit{Video1} of supplemental files). The ground truth is captured after 330ms of the input image. Here, the 330ms is approximating the round-trip delay considered by the PP node for the frame. Each frame of \textit{Video1} validates the performance, by comparing PP output with ground truth. Where, the ground truth is the real image captured after the delay considered in PP}. To analyze the role of this projection technique, some interesting areas are marked in the comparison. While performing the left turning manoeuvre, objects start moving right in the FOV. Some objects on the right, starts disappearing form the FOV. E.g. $Obj\,1$ marked is shifted towards right-bottom and $Obj\,2$ has completely disappeared from the FOV. This accurate transformation of the FOV, gives the human operator a feel of delay-free driving. Introduction of new objects on the left side of FOV can not be predicted (in case of just one camera facing forward), inpainting is used to fill the \textcolor{black}{null} pixels.

\begin{figure}
    \centering
    \subfloat[Trajectories comparison for left-angled R7-\ang{80} turn at 10 km/h]{%
        \includegraphics[width=\columnwidth]{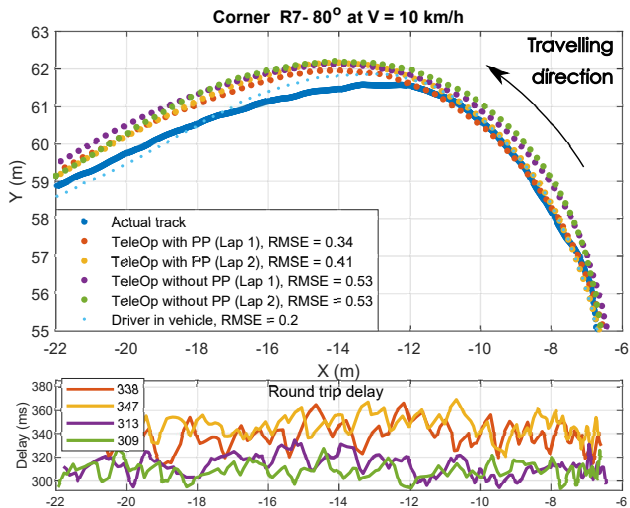}%
        \label{fig:a corner1}%
        }%
    \vfill%
    \subfloat[Trajectories comparison for left-angled R5-\ang{120} turn at 10 km/h]{%
        \includegraphics[width=\columnwidth]{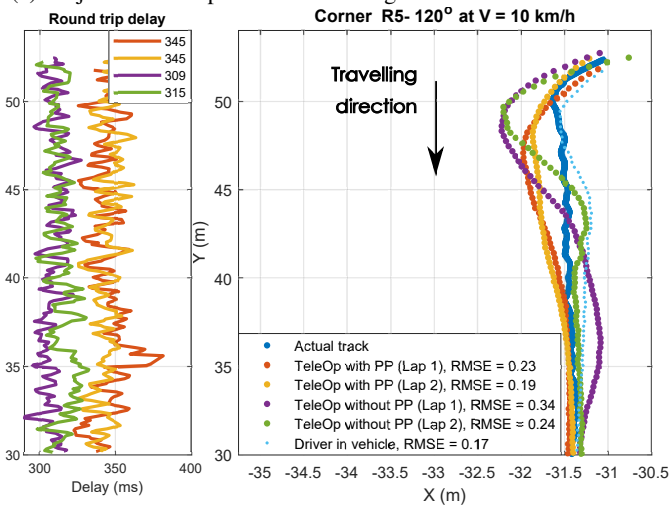}%
        \label{fig:b corner2}%
        }%
    \vfill%
    \subfloat[Trajectories comparison for double lane change at 10 km/h]{%
        \includegraphics[width=\columnwidth,cfbox=black .1pt .1pt]{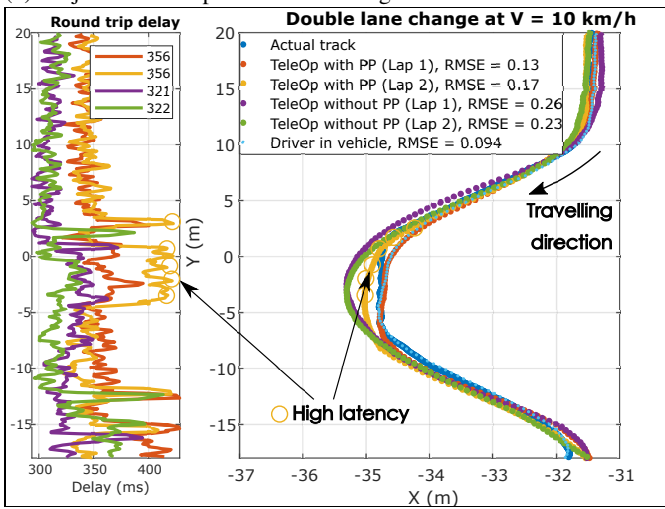}%
        \label{fig:c corner3}%
        }%
    \caption{\label{fig:y trajectories}Trajectories recorded in manoeuvres}
\end{figure}

Figure \ref{fig:a corner1} shows, trajectories traversed in all driving modes for the first R7-\ang{80}. Together with the vehicle trajectory, time delay experienced while driving at that particular X-coordinate is reported to correlate driving behaviour. This figure is stretched in y-axis to emphasise the cornering region. Deviation (\textcolor{black}{RMSE $0.2$ m}) is minimum in the case when the driver is inside the vehicle, reason is clear that there is no network delay between the driver and the vehicle. In case of teleoperation without projection, \textcolor{black}{deviation RMSE is $\sim0.53$m.} Performance improvement is observed in case of teleoperation with projection, where the \textcolor{black}{deviation RMSE is $\sim0.4$m} in both perspective projection (PP) laps. It is apparent in the figure that vehicle starts turning earlier ( at about $X=-12$m ) in case of perspective projection lap compare to the teleoperation lap without it. This is because projection tries to generate a perspective which the vehicle would observe, after traversing a distance corresponds to time delay in the network communication. It is also apparent that observed latency is consistently more in case of perspective projection. This is because projection algorithm requires RGB image as well as the depth-map. This means more computation, more bandwidth requirement, hence more time delay in data transmission. Interesting aspect is that performance is better even though latency is slightly increased (by $\sim30$ms).

Figure \ref{fig:b corner2} shows, trajectories traversed in all driving modes for the second corner R5-\ang{120}. Here, time delay experienced while driving at that particular Y-coordinate is reported. This figure is stretched in x-axis to emphasise the cornering region. This is a high curvature region of the test-track. In the first teleoperation lap without PP (purple), the vehicle starts turning after traversing a distance of around $0.5$m beyond the corner. To compensate this deviation, human operator steers more, due to which an oscillation is observed after the corner. This oscillation is induced due to the time-delay in the control loop between human and vehicle. In the second teleoperation lap without PP (green), the human operator tried to steer rapidly to be close to the track. But again a small oscillation is observed due to time-delay in the system. In the both laps of teleoperation with PP (red and green), \textcolor{black}{deviations are smaller} as it was in laps without PP. And, the trajectories are oscillation free, which is the advantage of smith predictor approach.

Figure \ref{fig:c corner3} shows, trajectories traversed for the double lane change manoeuvre. Here, time delay experienced while driving at that particular Y-coordinate is reported. This figure is also stretched in x-axis to emphasise the cornering regions. In the first teleoperation lap with PP (red), the deviation is quite less. In the second teleoperation lap with PP (yellow), the deviations are little greater than previous lap. This is because the latency in network was suddenly increased in that region, as it is apparent in the latency trend of yellow plot (marked with yellow circles between Y-coordinate -5 to 5). Even in increased latency region, deviation is not degraded compare to laps without PP.

\section{Discussion}
As seen in the results section, trajectories traversed with teleoperation without perspective projection outcomes a significant deviation and oscillations in narrow turns. This deviation is due to time-delay present in the information flow. Even for low vehicle speed of $10$ km/h, deviation of 0.5 meter is found for simple left-angled turn. This deviation may lead to unwanted incidents on street. Perspective projection technique doesn't completely eliminate the deviations, but reduces the deviations and eliminates oscillations in the trajectory. Even the time-delay in the loop is more in case of perspective projection, trajectories found to be stable and close to the reference as compare to that of without the perspective projection.
\\\\
\emph{Limitations}: As the vehicle speed increases ($>10$ km/h), predictive image becomes blurry because of noise in depth-map. Also, it does not take into account the independent motion of objects in the frame. It considers objects to be still, and just the vehicle in motion. For small motion of objects and for small time-delay, this assumption is reasonable. 
\\\\
\emph{Advantages}: Performing tele-operation with perspective projection resulted in significant reduction in deviations while maneuvering for corner R7-\ang{80} and double lane change. In high curvature manoeuvre of corner R5-\ang{120} at $10$ km/h, oscillations are eliminated as compared with teleoperation laps without perspective projection.
\section{Conclusion}
To mitigate the detrimental effect of time-delay in vehicle teleoperation, perspective projection technique is used to transform the image streaming. Smith predictor approach is used to estimate the correction which can be added to the inputs (images) given to the human operator. The plant model inside the smith predictor is a single track kinematic vehicle model, which is a reasonable simplification considering low speed in vehicle teleoperation. Perspective projection technique merges the correction given by smith predictor to the streaming images. It is able to generate the transformed image not only in straight driving condition but also during cornering condition. One of its input is the depth-map. In order to transmit depth-map to the control-station, it is first transformed into 8 bit image using a mathematical relation which considers linear increasing resolution for depth measurements. Both, RGB and depth images are transmitted to the control-station in JPEG. For estimation of uplink-delay, stochastic approach is used to estimate $95^{th}$ percentile of the latest trend of uplink-delays in the network. The human-operator sees a perspective projected image of the vehicle surrounding. This display tries to emulate vehicle's perspective when the vehicle would receive driving commands. Due to this, human-operator is able to make driving decision bit in advance. Which results, better control in following desired trajectory.

For validation of the approach, vehicle teleoperation is performed on some edge-case scenarios of street driving. Scenarios are cornering and double-lane change at $10$ km/h. To exhibit the improvements, results of two laps with and without PP are compared with each other. Table \ref{tab:summaryTable} presents the RMS deviation during the manoeuvres. With PP, the deviations and oscillations are found less as compared to vehicle teleoperation without it, which improves the confidence of human operator. 
\begin{table}[H]
\begin{tabular}{|l|l|l|l|l|
>{\columncolor[HTML]{9DE7FF}}l |}
\hline
                                                                           & \multicolumn{2}{l|}{TeleOp with PP}                                                                   & \multicolumn{2}{l|}{TeleOp without PP}                                                                & \cellcolor[HTML]{9DE7FF}                                                                              \\ \cline{2-5}
\multirow{-2}{*}{\begin{tabular}[c]{@{}l@{}}RMS\\ Deviations (m)\end{tabular}} & \multicolumn{1}{c|}{\cellcolor[HTML]{FFCCC9}Lap1} & \multicolumn{1}{c|}{\cellcolor[HTML]{FFFFC7}Lap2} & \multicolumn{1}{c|}{\cellcolor[HTML]{CBCEFB}Lap1} & \multicolumn{1}{c|}{\cellcolor[HTML]{9AFF99}Lap2} & \multirow{-2}{*}{\cellcolor[HTML]{9DE7FF}\begin{tabular}[c]{@{}l@{}}Driver in\\ vehicle\end{tabular}} \\ \hline
R7-\ang{80}                                                                      & \cellcolor[HTML]{FFCCC9}0.34                      & \cellcolor[HTML]{FFFFC7}0.41                      & \cellcolor[HTML]{CBCEFB}0.53                      & \cellcolor[HTML]{9AFF99}0.53                      & 0.20                                                                                                  \\ \hline
R5-\ang{120}                                                                     & \cellcolor[HTML]{FFCCC9}0.23                      & \cellcolor[HTML]{FFFFC7}0.19                      & \cellcolor[HTML]{CBCEFB}0.34                      & \cellcolor[HTML]{9AFF99}0.24                      & 0.17                                                                                                  \\ \hline
\begin{tabular}[c]{@{}l@{}}Double lane\\ change\end{tabular}               & \cellcolor[HTML]{FFCCC9}0.20                      & \cellcolor[HTML]{FFFFC7}0.21                      & \cellcolor[HTML]{CBCEFB}0.26                      & \cellcolor[HTML]{9AFF99}0.25                      & 0.11                                                                                                  \\ \hline
\end{tabular}
\caption{Deviations recorded in manoeuvres with and without perspective projection.}
\label{tab:summaryTable}
\end{table}
\textcolor{black}{\emph{Future work:} Multi cameras and dynamic vehicle model (for position prediction) which would also require a vehicle state estimator to transmit vehicle states in addition to the images. Further assessment of vehicle teleoperation performance by incorporating trajectory and drive feeling result of more voluntary human operators.}
\section*{Acknowledgements}
This work is supported by \emph{FESR 2014-2020 PROGETTO ID 242092 - TEINVEIN (TECNOLOGIE INNOVATIVE PER I VEICOLI INTELLIGENTI) CUP E96D17000110009}

\ifCLASSOPTIONcaptionsoff
  \newpage
\fi

\bibliographystyle{IEEEtran}
\bibliography{sample}



\end{document}